\documentclass[showpacs,preprintnumbers,
amsmath,amssymb,groupedaddress,superscriptaddress]{revtex4}
\usepackage{graphicx}
\usepackage{dcolumn}
\usepackage{bm}

\begin{document}
\draft \title{Theoretical mean field and experimental occupation
  probabilities in the double beta decay system $^{76}$Ge to
  $^{76}$Se}

\author{O. Moreno}
\address{Dpto. F\'isica At\'omica, Molecular y Nuclear, Univ. Complutense de Madrid, E-28040 Madrid, Spain}
\author{E. Moya de Guerra}
\address{Dpto. F\'isica At\'omica, Molecular y Nuclear, Univ. Complutense de Madrid, E-28040 Madrid, Spain}
\author{P. Sarriguren}
\address{Instituto de Estructura de la Materia, CSIC, Serrano 123, E-28006 Madrid, Spain}
\author{Amand Faessler}
\address{Institut f\"ur Theoretische Physik, Universit\"at T\"ubingen, D-72076 T\"ubingen, Germany}

\date{\today}

\begin{abstract}
  Usual Woods-Saxon single particle levels with BCS pairing are not
  able to reproduce the experimental occupation probabilities of the
  proton and neutron levels $1p_{3/2}$, $1p_{1/2}$, $0f_{5/2}$,
  $0g_{9/2}$ in the double-beta decay system $^{76}$Ge to $^{76}$Se.
  Shifting down the $0g_{9/2}$ level by hand can explain the data but
  it is not satisfactory. Here it is shown that a selfconsistent
  Hartree-Fock+BCS approach with experimental deformations for
  $^{76}$Ge and $^{76}$Se may decisively improve the agreement with
  the recent data on occupation probabilities by Schiffer {\it et al.}
  and Kay {\it et al.} Best agreement with available data on $^{76}$Ge
  and $^{76}$Se, as well as on neighbor isotopes, is obtained when the
  spin-orbit strength for neutrons is allowed to be larger than that
  for protons. The two-neutrino double-beta decay matrix element is
  also shown to agree with data.

\end{abstract}


\maketitle

Among double-beta decay with two neutrino processes, the best studied
case is that of $^{76}$Ge going to $^{76}$Se. Within a spherically
symmetric description of these nuclei, the relevant proton and neutron
single particle levels involved in the double-beta decay process are
in the valence shells $1p_{3/2}$, $1p_{1/2}$, $0f_{5/2}$,
$0g_{9/2}$. Their occupation probabilities have been recently measured
by Schiffer {\it et al.}  \cite{sch08} (for neutrons) and by Kay {\it
  et al.}  \cite{kay09} (for protons). These data are not reproduced
by previous mean field calculations using Woods-Saxon potential and
pairing within BCS approximation.  By increasing the binding energy of
the $0g_{9/2}$ level the agreement is better, since its occupation
probability increases and eventually reaches the experimental
one. This energy shifting has been performed by hand in previous works
\cite{suh08, sim09, bar09} to obtain the experimental occupation
probabilities, and the resulting ground state structure has been used
to compute neutrinoless double-beta matrix elements.  Shell-model
calculations \cite{men09} have also been carried out with occupation
probabilities in good agreement with experimental ones, but the single
particle energies used are not available in the literature.

In this work we use a deformed Skyrme Hartree-Fock (HF) mean field
with pairing correlations within BCS approximation. At each HF
iteration the BCS equations are solved to get new single particle wave
functions $\Phi_i$, energies $e_i$ and occupation probabilities
$v_i^2$ until convergence is achieved. After convergence, the
proton-neutron quasiparticle random phase approximation (pnQRPA)
equations are solved on each deformed ground state basis for $^{76}$Ge
and $^{76}$Se to get their Gamow-Teller (GT) strength distributions
and to compute the two-neutrino double-beta decay matrix element. The
formalism is similar to that used in previous works
\cite{sar03,alv04,mor09} with an important difference. In this work we
allow the Skyrme force parameters to vary and search for a
parametrization that provides better agreement with the experimental
occupations of the above mentioned valence shells in $^{76}$Ge and
$^{76}$Se, as well as in their neighbors. We focus on the effect of
the spin-orbit strength, $W_0$, that we find to be of primary
importance for the desired description of proton and neutron valence
shells occupations.

\begin{figure*}
\centering \includegraphics[width=180mm]{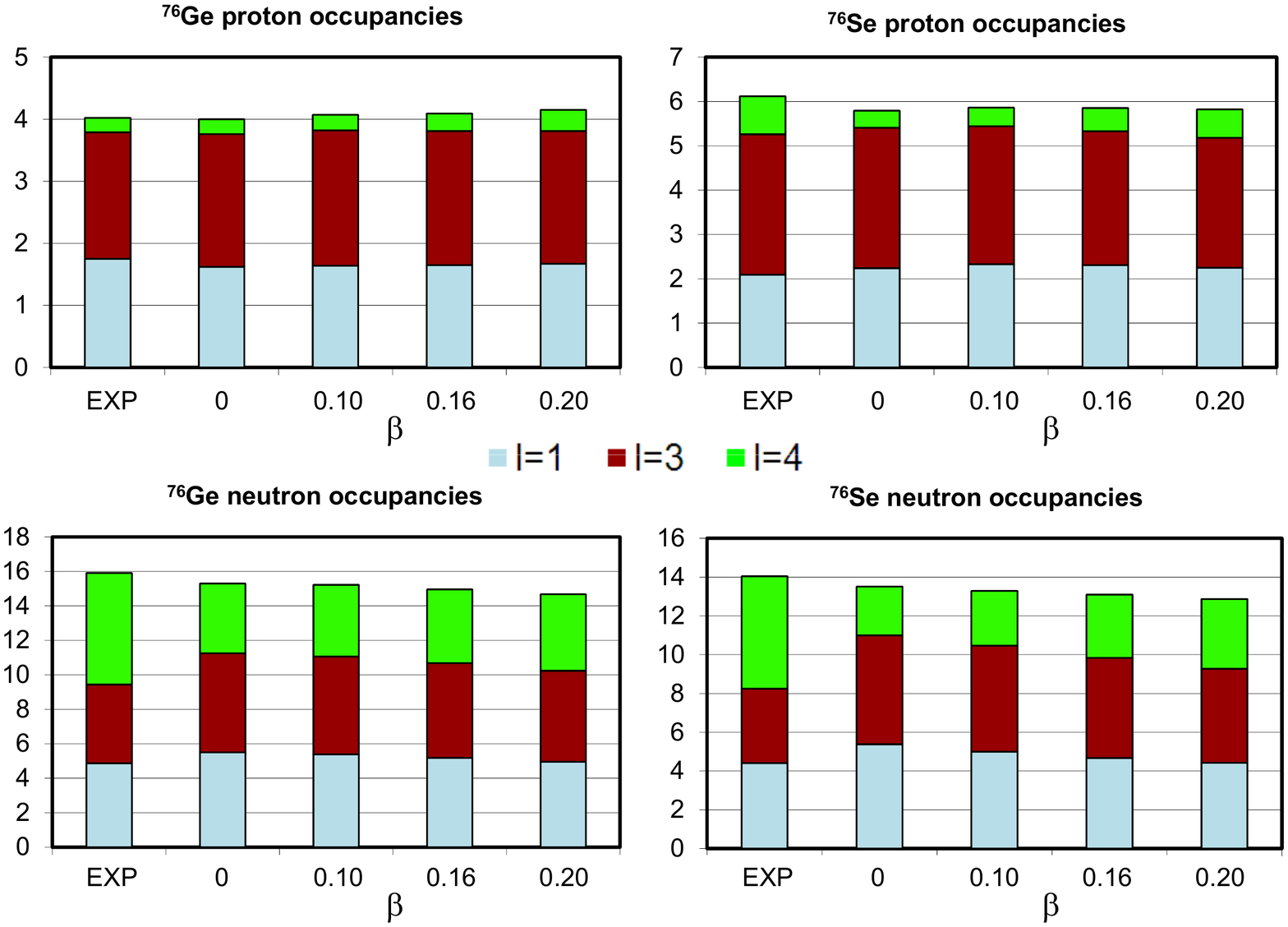}
\caption[]{(Color online) Active-shell experimental \cite{sch08,kay09}
  and theoretical occupations for different quadrupole deformations
  (indicated by the deformation parameter $\beta$), from a HF(Sk3)+BCS
  calculation in $^{76}$Ge and $^{76}$Se. Occupations are shown for
  the proton and neutron states in the active shells, which include
  the single particle states $1p_{1/2}$ and $1p_{3/2}$ (gathered as
  $l=1$), $0f_{5/2}$ ($l=3$) and $0g_{9/2}$ ($l=4$).}
\label{fig_occup_def}
\end{figure*}

Experimental data and HF+BCS calculations suggets prolate deformations
for the $^{76}$Ge and $^{76}$Se ground states. The most precise data
on the quadrupole moments of these nuclei come from Coulomb excitation
reorientation measurements \cite{rag89} which, together with
experimental charge radii \cite{ang04}, give a deformation parameter
$\beta=$0.10 for $^{76}$Ge and $\beta=$0.16 for $^{76}$Se (with
25-30$\%$ uncertainty). These experimental deformations are taken into
account when obtaining our theoretical occupation probabilities in the
active shell, and naturally leads us to check if the quadrupole
deformation is responsible (fully or in part) for the disagreement
between previous theoretical occupations and the experimental ones. To
analyze this we show in Fig.~\ref{fig_occup_def} the experimental
occupations in the active shell together with our theoretical results
for different values of the quadrupole deformation, including the
experimental and selfconsistent one, constraining the quadrupole
moment in the HF+BCS procedure with Sk3 Skyrme force
\cite{bei75}. Since the single-particle orbital and total angular
momenta are not good quantum numbers in the deformed cases, a
projection to a spherical basis has been performed to obtain the
occupation probabilities of the spherical valence shells.  The
uncertainties of the experimental data, not shown in the plot, are of
a few percent (less than 10$\%$), except for the highly unoccupied
$l=4$ proton state whose uncertainty is 30$\%$ in Se and over 100$\%$
in Ge.  As the quadrupole deformation increases, the trend that can be
observed in the figure is a decrease in the occupation probabilities
of the states $l=1$ ($1p_{3/2}$ and $1p_{1/2}$ together) and $l=3$
($0f_{5/2}$) and an increase in the occupation of the state $l=4$
($0g_{9/2}$).  Although this effect goes in the right direction, there
are some features of the experimental occupations that are not
reproduced by the deformed calculations as long as the deformation is
constrained to reasonable values agreeing with experiment. The most
noticeable failure of the theory is a lack of occupancy in the
$0g_{9/2}$ neutron level. We have checked that this failure is shared
by a large variety of the Skyrme forces where the Skyrme spin-orbit
strength $W_0$ is equal for protons and neutrons. In any case, the
selfconsistent deformed single nucleon potential gives better
agreement with experiment for the occupation probabilities than a
Woods-Saxon potential \cite{suh08,sim09}.

To improve agreement with experimental occupations, we found it
crucial to vary the spin-orbit terms of the Skyrme energy density
functional. More specifically, to increase the value of the $W_0$
parameter in the HF+BCS potential for neutrons.

\begin{figure*}
\centering \includegraphics[width=180mm]{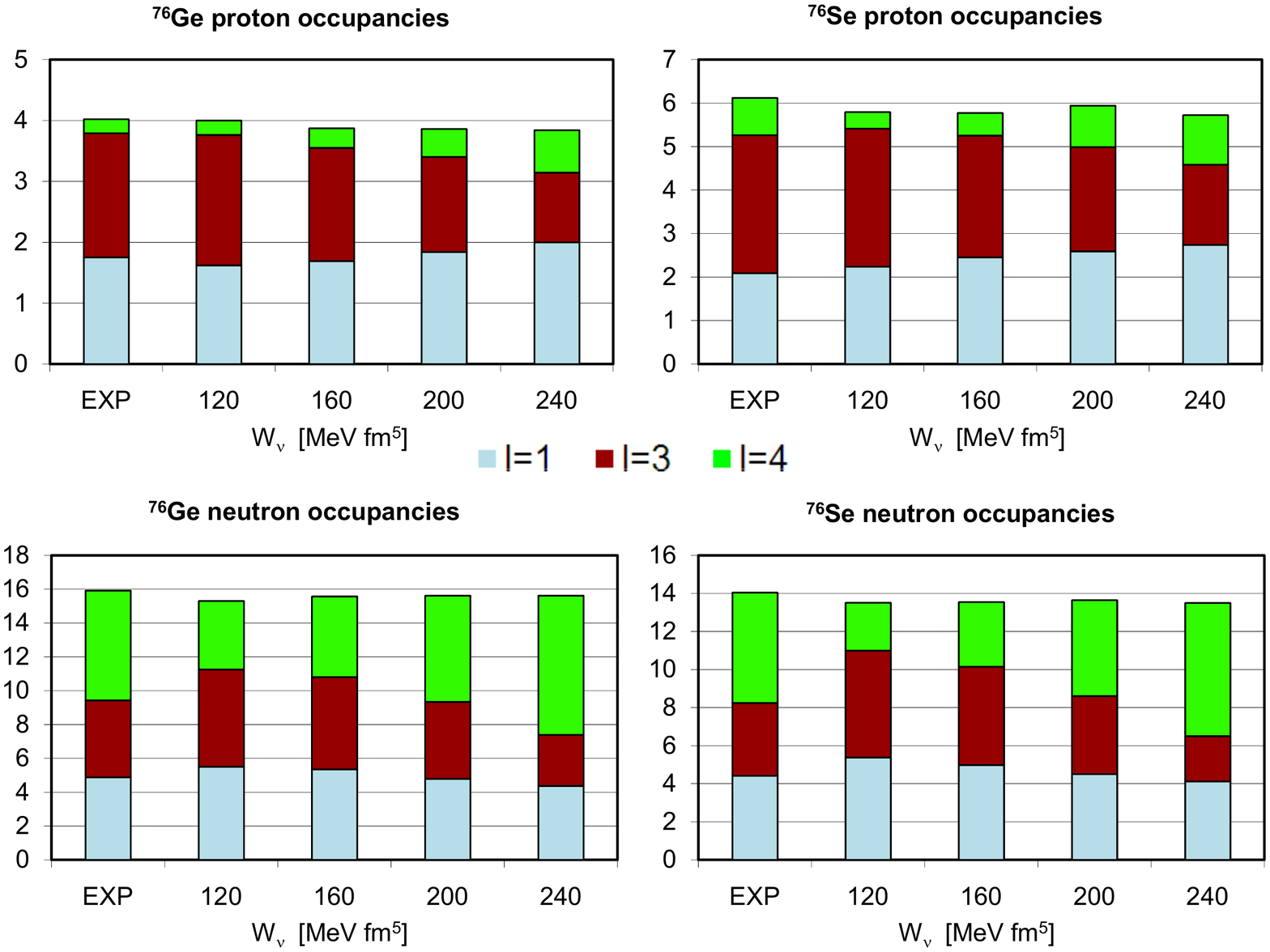}
\caption[]{(Color online) Active-shell experimental \cite{sch08,kay09} and
  theoretical occupations for different values of the neutron
  spin-orbit strength $W_{\nu}$ (with $W_{\pi}=$120), from a
  HF(Sk3$^{\nu}$)+BCS calculation in $^{76}$Ge and $^{76}$Se with
  spherical ground states.}
\label{fig_occup_Wn}
\end{figure*}

It has been already pointed out in the past that the Skyrme
interactions do not reproduce properly the spin-orbit splittings of
some nuclei, especially concerning neutron states, and that a more
flexible two-body spin-orbit interaction is required \cite{rei95}. The
effect of modifying the Sk3 spin-orbit parameter $W_0$ for neutrons is
shown in Fig.~\ref{fig_occup_Wn}, which contains results from a
HF(Sk3$^{\nu}$)+BCS calculation for spherical shape. The generalized
Skyrme interaction Sk3$^{\nu}$ maintains the original Sk3
parametrization except for the value of the spin-orbit strength $W_0$
for neutrons that has been replaced by a new strength $W_{\nu}$
affecting only the terms of the potentials containing neutron
densities. The value of the neutron strength has been modified from an
initial value $W_{\nu}=W_0=$120 MeV fm$^5$ all the way up to the value
$W_{\nu}=$240 MeV fm$^5$. The corresponding strength for protons has
been kept fixed, $W_{\pi}=W_0=$120 MeV fm$^5$ (the units of this
parameter will be omitted in the following discussions). It is worth
pointing out that because of selfconsistency the change of $W_0$ for
neutrons affects not only the neutron levels but also de proton
levels.

As shown in the plot, a good agreement on neutron occupation
probabilities between theory and experiment is found for
$W_{\nu}=$200; in this situation the $0g_{9/2}$ state is more bound
than the $1p_{1/2}$ state. The result for protons yields also a good
agreement with experiment; in this case, the proton state $1p_{1/2}$
is still more bound than the state $0g_{9/2}$.

\begin{figure*}
\centering \includegraphics[width=180mm]{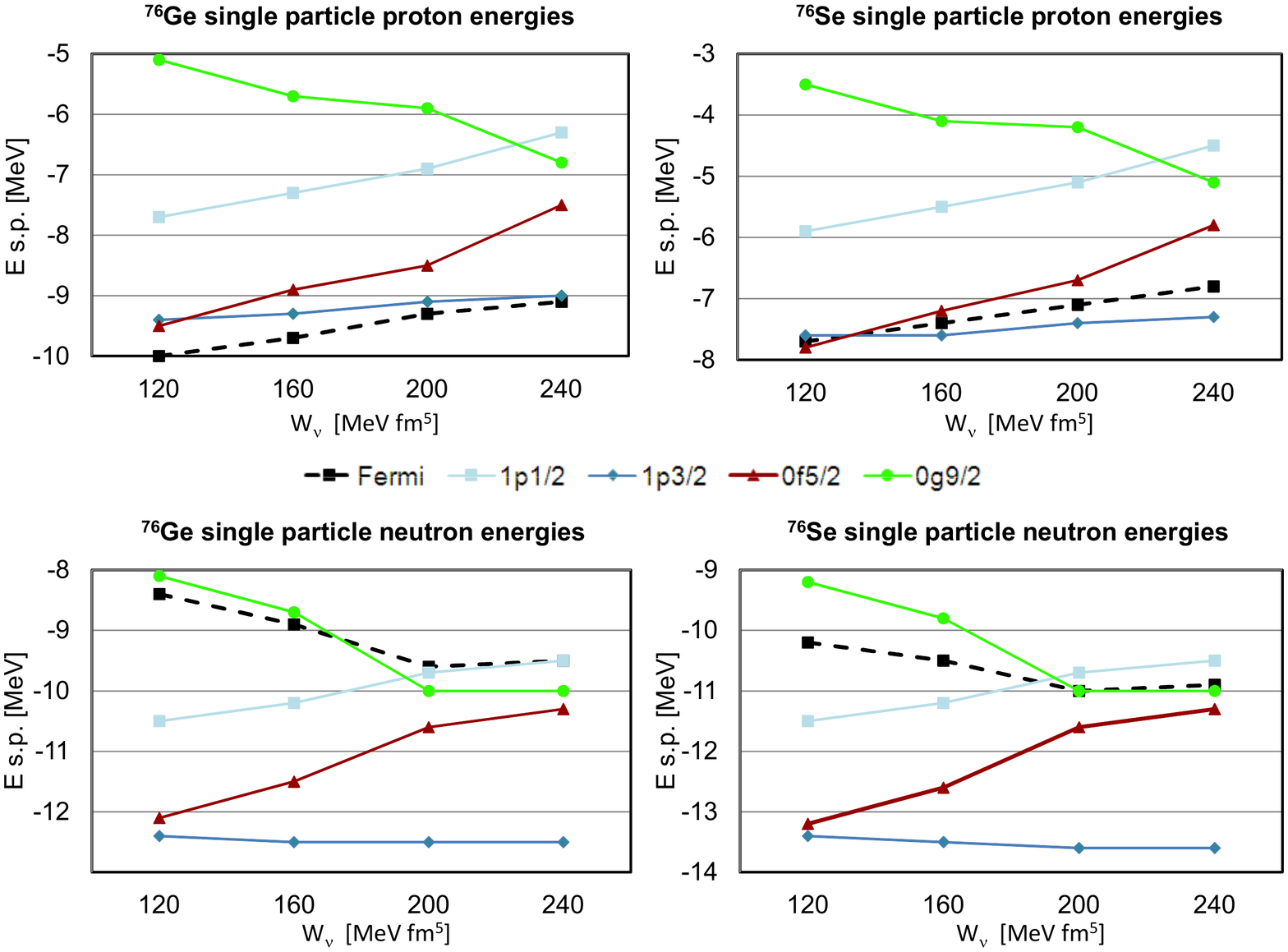}
\caption[]{(Color online) Theoretical Fermi levels and active-shell
  single particle energies for different values of the neutron
  spin-orbit strength $W_{\nu}$ (with $W_{\pi}=$120), from a
  HF(Sk3$^{\nu}$)+BCS calculation in $^{76}$Ge and $^{76}$Se with
  spherical ground states. The occupations corresponding to these
  levels appeared in Fig.~\ref{fig_occup_Wn}.}
\label{fig_spenergies}
\end{figure*}

In Fig.~\ref{fig_spenergies} one can see the proton and neutron single
particle energies in the active shell corresponding to the occupation
probabilities of Fig.~\ref{fig_occup_Wn}, together with the Fermi
levels. As the spin-orbit strength for neutrons increases, the single
proton and neutron levels $1p_{1/2}$ and $0f_{5/2}$ get clearly less
bound, whereas the level $0g_{9/2}$ gets more bound. The level
$0g_{9/2}$ exchanges its energy position with the level $1p_{1/2}$ in
the vicinity of $W_{\nu}=$240 in the case of protons and for
160$<W_{\nu}<$200 in the case of neutrons. In this last case, the
level $0g_{9/2}$ also exchanges its energy position with the level
$0f_{5/2}$ but in the range 200$<W_{\nu}<$240. Finally, the level
$1p_{3/2}$, which is in general the most bound of them, remains with
about the same binding energy. In what follows we refer to the
interaction with $W_{\pi}=$ 120 and $W_{\nu}=$ 200 as Sk3$^{\nu}$.

The considerable improvement obtained by setting different values of
the spin-orbit interaction for each type of nucleon supports a
modification of the spin-orbit term in the general Skyrme interaction
\cite{vau72}, that could have a structure similar to that of the other
Skyrme terms, namely containing a factor of the form $(1+\epsilon
P_{\tau})$, where $P_{\tau}$ is an isospin-exchange operator. More
involved isospin-dependent formulations are also possible and worth
exploring.

\begin{figure*}
\centering \includegraphics[width=180mm]{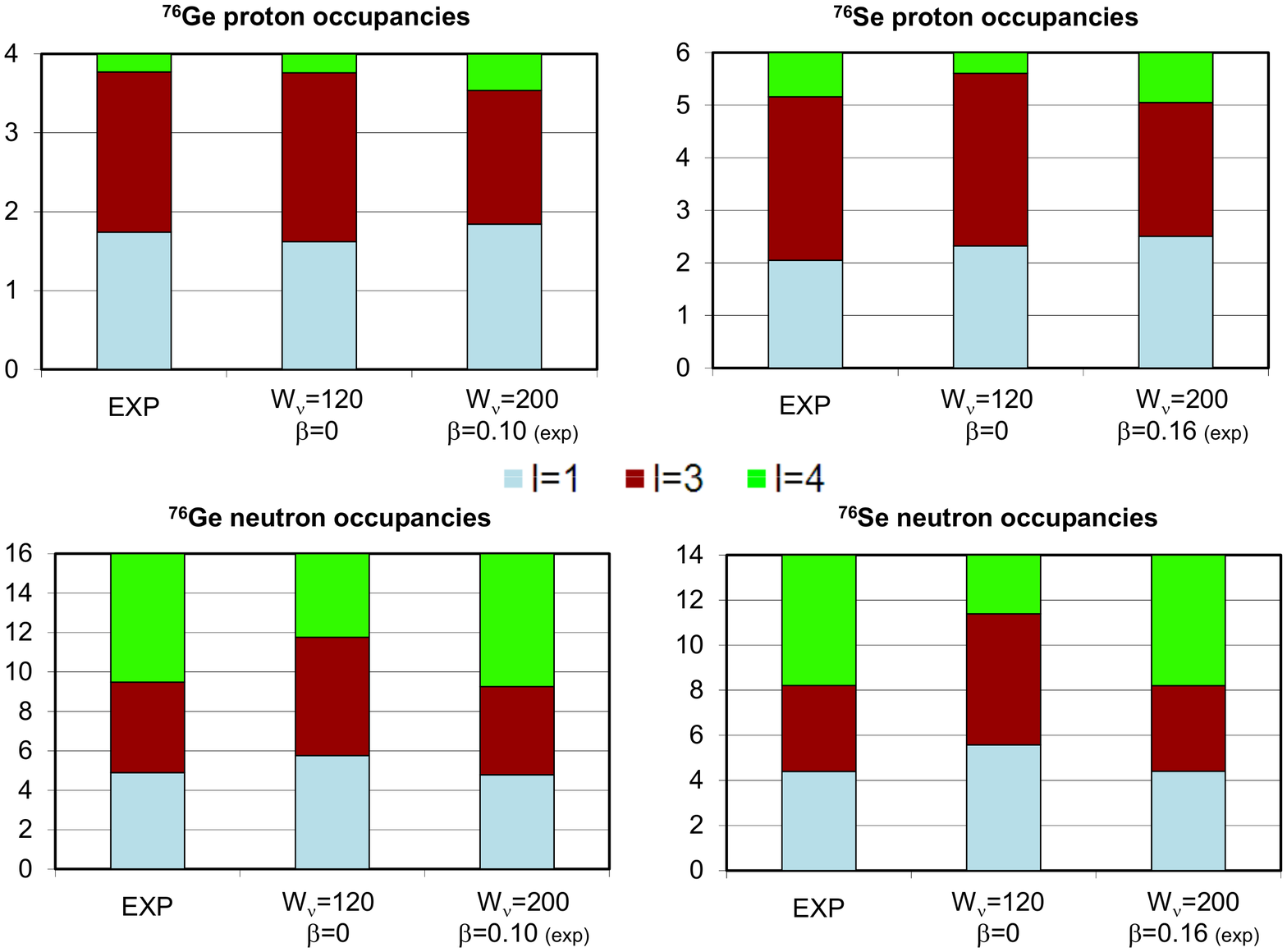}
\caption[]{(Color online) Active-shell experimental \cite{sch08,kay09}
  and theoretical occupations in $^{76}$Ge and $^{76}$Se. The first
  theoretical result is the original HF(Sk3)+BCS calculation for
  spherical ground states and the second one is the HF(Sk3$^{\nu}$)+BCS
  calculation with $W_{\nu}=$200 and experimental ground-state
  deformations \cite{rag89}.  Both experiment and theory have been
  normalized so that the active shell contains all the nucleons
  exceeding the magic number 28.}
\label{fig_occup_final_ge76se76}
\end{figure*}

We show in Fig.~\ref{fig_occup_final_ge76se76} the occupation
probabilities in the active shells obtained with spherical HF(Sk3)+BCS
and with the improved HF(Sk3$^{\nu}$)+BCS (using $W_{\pi}=$120 and
$W_{\nu}=$200) and the experimental deformation.  The results are also
compared to the experimental data. All of them are normalized so that
the total number of nucleons in the active shells is the expected one:
4 protons and 16 neutrons in $^{76}$Ge and 6 protons and 14 neutrons
in $^{76}$Se. In our calculations the actual numbers of protons and
neutrons in these valence shells are lower, especially in the deformed
cases, where the occupation probabilities are spread along many
spherical single particle states, not only in those active shells of
the spherical shell model. When obtaining the occupations from
experiment, a similar (although not identical) renormalization
procedure was also performed (see \cite{sch08, kay09} for details). As
seen in the figure, the improvement reached by combining the effect of
deformation and modified spin-orbit strength with respect to the
spherical HF(Sk3$^{\nu}$)+BCS is apparent. It is possible however that
other ingredients are still missing which could enhance the trend
initiated by deformation and spin-orbit strength towards a reliable
description of the single particle spectrum in this nuclear region.

\begin{figure*}
\centering \includegraphics[width=180mm]{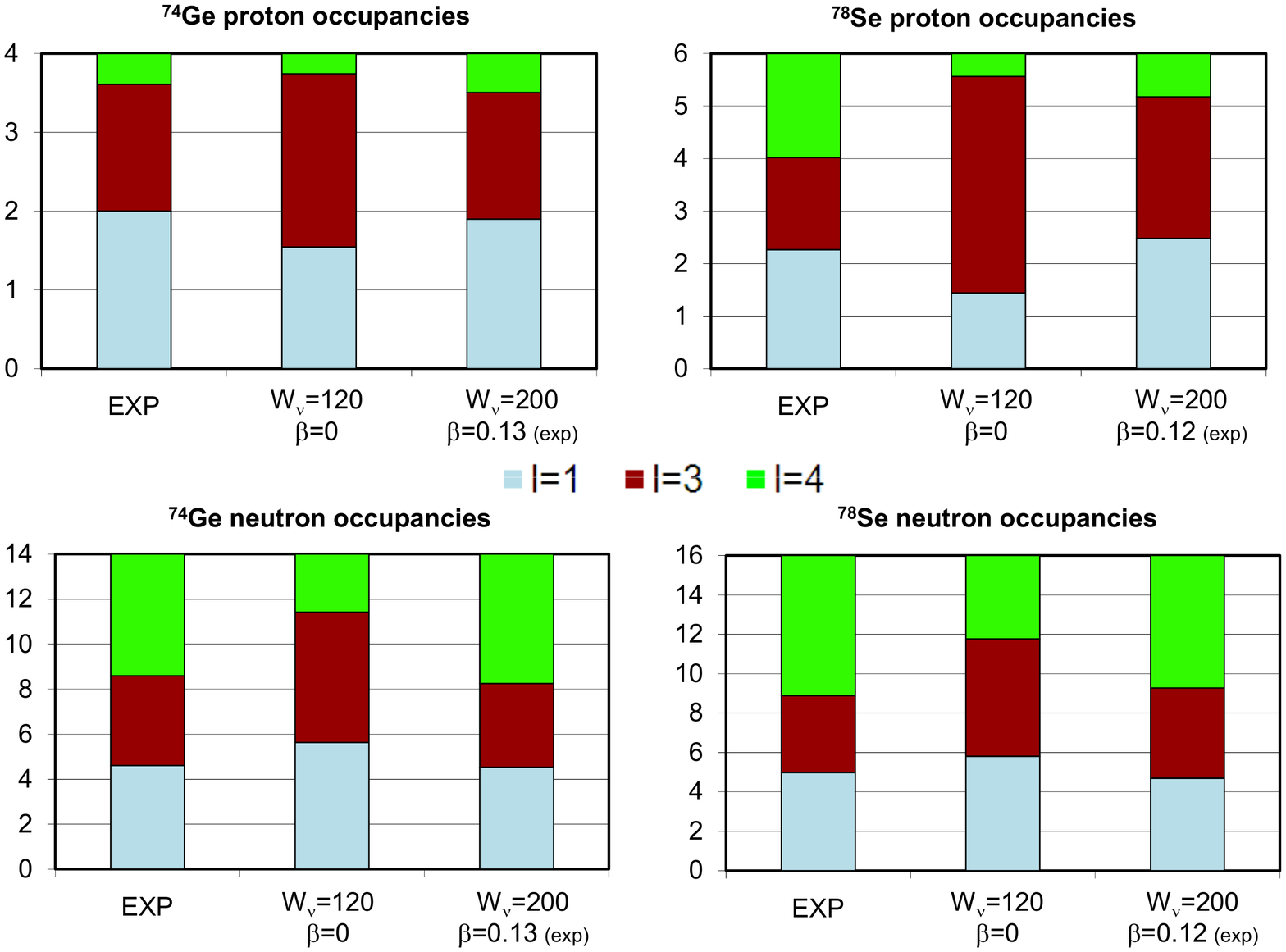}
\caption[]{(Color online) Same as in
  Fig.~\ref{fig_occup_final_ge76se76} but for $^{74}$Ge and
  $^{78}$Se.}
\label{fig_occup_final_ge74se78}
\end{figure*}

This new spin-orbit strength can be applied to the neighbor nuclei
$^{74}$Ge and $^{78}$Se, whose active-shell occupation probabilities
have been also measured \cite{sch08,kay09}. In
Fig.~\ref{fig_occup_final_ge74se78} we compare the measured
probabilities with our theoretical results for the HF(Sk3)+BCS
calculation with spherical shape and for the new HF(Sk3$^*$)+BCS
calculation with experimental deformations. The $\beta$-values
extracted from data in refs.~\cite{rag89,ang04} are $\beta=$ 0.13 for
$^{74}$Ge and $\beta=$ 0.12 for $^{78}$Se. The inclusion of
experimental deformation and a larger spin-orbit strength for neutrons
change also in this case the theoretical results towards a better
agreement with the experimental data.

\begin{figure*}
\centering \includegraphics[width=180mm]{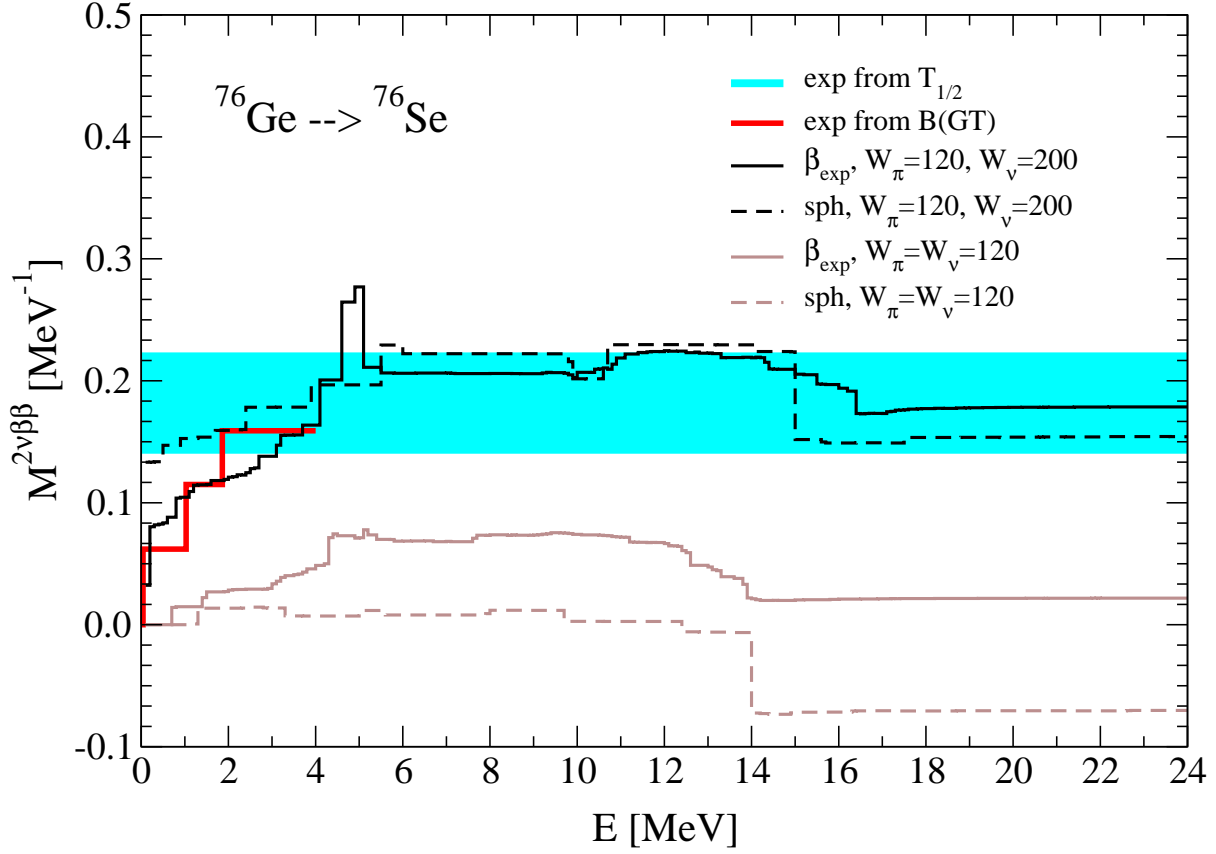}
\caption[]{(Color online) Running sum of the matrix element for the
  two-neutrino double-beta decay $^{76}$Ge$\to$$^{76}$Se, as a
  function of the intermediate excitation energy in $^{76}$As. The
  single particle basis for $^{76}$Ge and $^{76}$Se have been obtained
  from HF(Sk3$^{\nu}$)+BCS ($W_{\nu}>W_{\pi}$) and from HF(Sk3)+BCS
  ($W_{\nu}=W_{\pi}=W_0$) calculations for spherical (sph) and
  experimental deformed (exp def) ground state shapes. An
  experimental range of the total matrix element is shown, coming from
  the experimental half-life of the process.  The experimental
  low-energy steps of the matrix element running sum come from
  experimental GT$^+$ and GT$^-$ strengths \cite{gre08}.}
\label{2b_gese_SO_def}
\end{figure*}

The new single particle energies and occupation probabilities have
also an effect on the matrix element of the two-neutrino double-beta
decay $^{76}$Ge$\to$$^{76}$Se. In Fig.~\ref{2b_gese_SO_def} we show
the theoretical double-beta decay matrix element running sum as a
function of the excitation energy of the intermediate nucleus
$^{76}$As.  The results shown correspond to deformed pnQRPA
calculations (see ref.~\cite{alv04} and references therein for details
of calculations) using the single particle basis for $^{76}$Ge and for
$^{76}$Se obtained with the new (Sk3$^{\nu}$, $W_{\nu}>W_{\pi}$) and
the standard (Sk3, $W_{\nu}=W_{\pi}=W_0$) parametrizations, with and
without deformation. Proton-neutron particle-hole $ph$ and
particle-particle $pp$ residual interactions within QRPA are included
to obtain the contributions to the double-beta matrix element,
following the procedure explained in previous works \cite{alv04}. The
coupling constants of the $ph$ and $pp$ channels take the values
$\chi_{ph}=$ 0.25 MeV and $\kappa_{pp}=$ 0.08 MeV respectively. In the
case of $\chi_{ph}$, it is an average value between the one used in
previous works \cite{mor09,hom96,sal09} and the selfconsistent value
\cite{sar98}. In the case of $\kappa_{pp}$, the current value is in
agreement with the one used in some previous works \cite{alv04,sal09},
but it is twice as large as the one in \cite{mor09}. We note that the
larger value of the $pp$ residual interaction coupling constant is
responsible for the particular shape of the plots in
Fig.~\ref{2b_gese_SO_def}, where the matrix element increases in the
region from 0 to 4 MeV of excitation energy, then goes up and down
with increasing energy and finally decreases until reaching its final
value. This behavior does not show up for the smaller values of the
$pp$ coupling constant in \cite{mor09}. This finding also agrees with
the conclusions reached in ref.~\cite{fan09}, namely that the
characteristic increase and decrease with energy of the matrix element
found in many shell model calculations is also found in our deformed
pnQRPA calculations when we use larger values of $\kappa_{pp}$ than
those used in our previous works \cite{mor09}.

The experimental data shown in Fig.~\ref{2b_gese_SO_def} are obtained
in a phenomenological way using the experimental Gamow-Teller (GT)
strengths of the transitions $^{76}$Ge$\to$$^{76}$As (GT$^{-}$)
\cite{mad89} and $^{76}$Se$\to$$^{76}$As (GT$^{+}$) \cite{gre08}, as
explained in previous experimental or theoretical works
\cite{gre08,mor09}. An experimental range for the final value of the
matrix element is also shown in the plot, obtained from the
experimental double-beta decay half-life \cite{bar06} with two
possible values of the axial-to-vector ratio $g_A=1.25$ and $g_A=1.00$
(quenched value).

It is clearly seen in the figure that the calculations with the
increased spin-orbit strength for neutrons reproduce much better the
experimental values. In the low energy region the agreement is
improved with deformation. There are no large differences between the
calculations with spherical ground states and the one with
experimental deformations. For Sk3$^{\nu}$ both of them lie within the
experimental range for the total value of the matrix element. For Sk3
a large disagreement is found with the $\kappa_{pp}$ value used here.

We can trace back the origin of the low-energy behavior of the
double-beta matrix element, studying the accumulated strength
distributions of the GT$^-$ transition $^{76}$Ge$\to$$^{76}$As and of
the GT$^+$ transition $^{76}$Se$\to$$^{76}$As. In the case of the
GT$^-$ strength distribution, our result with $W_{\nu}>W_{\pi}$ is
roughly twice as large as the result with the usual strength
($W_{\nu}=W_{\pi}=W_0$) until 4 MeV, and the former reproduces better
the experimental data in this range. In the case of the GT$^+$
strength distribution, the results with increased spin-orbit strength
are again larger, and there is a big difference between the spherical
or the deformed single particle basis for $^{76}$Se. The deformed one
gives the best agreement with the results using experimental
data. Beyond 2 MeV the experimental value increases, whereas the
theoretical results wait until 5 MeV to do so (not shown in the plot).

From this analysis we conclude that the better agreement between
theoretical and experimental occupation probabilities in the valence
states with deformation and enhanced neutron spin-orbit gives rise to
GT transitions in good agreement with the experimentally known
strengths at low energy that involve the valence shells. This in turn
produce a better agreement for the double-beta matrix element.

Apart from the single- and double-beta decay properties of the nuclei
under study, the modification of the neutron spin-orbit strength
affects some bulk properties of the nuclei, in particular the binding
energy and the charge radius. Since the parameters of the Skyrme
interactions are fitted to reproduce these bulk properties in a given
set of nuclei, the modification of one of them without an overall
refitting of the rest may result in unrealistic values of those bulk
properties. Notwithstanding the above, when comparing with the
experimental binding energies of 661.6 MeV for $^{76}$Ge and 662.07
MeV for $^{76}$Se we get a 0.5$\%$ underestimation in the calculation
with the usual Sk3 force and a 2$\%$ overestimation in the calculation
with the Sk3$^{\nu}$ force. Concerning the charge radii, the
HF(Sk3)+BCS results are 4.12 fm for $^{76}$Ge and 4.17 fm for
$^{76}$Se, whereas with increased spin-orbit strength one gets 4.09 fm
for $^{76}$Ge and 4.14 fm for $^{76}$Se, which actually agree better
with the experimental values of 4.08 fm for $^{76}$Ge and of 4.14 fm
for $^{76}$Se \cite{ang04}.

To conclude, we have shown that agreement with experimental occupation
numbers in the valence shells of $^{76}$Ge and $^{76}$Se can be
obtained within the selfconsistent Skyrme+HF+BCS deformed mean field
approximation. Best agreement with experiment is obtained when a
spin-orbit interaction more flexible than the standard one in Skyrme
forces is considered, allowing in particular for a larger neutron
spin-orbit force. We point out that the interplay between spin-orbit
strengths and deformation is an essential feature to give the
experimental occupations. These occupation numbers are in turn of
great importance for beta strength distributions and two-neutrino
double-beta decay matrix element calculations, especially in the low
energy region. Changes at the level of the two-body Skyrme interaction
and energy density functional are suggested to further improve the
theoretical description of nuclei and will be further investigated.

We acknowledge the support of the Internationales Graduiertenkolleg
GRK683 of the DFG and the Sonderforschungsbereich TR27, Ministerio de
Ciencia e Innovaci\'on (Spain) under Contract. Nos.  FIS2008-01301 and
CSPD-2007-00042@Ingenio2010; and CSIC-UCM `Grupo de F\'isica Nuclear'
- 910059 and FPA-2007-62616.

\newpage

\end{document}